\documentclass[a4paper,12pt]{article}
\usepackage{graphicx,rotating,colortbl,xcolor,wrapfig,hyperref,slashed,cite,multicol}

\hypersetup{colorlinks,bookmarksopen,bookmarksnumbered,
linkcolor=blus,pdfstartview=FitH,urlcolor=rossos,citecolor=verde}

\def\to{\rightarrow}
\def\bea{\begin{eqnarray}}
\def\eea{\end{eqnarray}}
\def\eq#1{~(\ref{eq:#1})}

\definecolor{Gray}{gray}{0.95}

\definecolor{rosso}{cmyk}{0,1,1,0.4}
\definecolor{rossos}{cmyk}{0,1,1,0.55}
\definecolor{rossoc}{cmyk}{0,1,1,0.2}
\definecolor{blu}{cmyk}{1,1,0,0.3}
\definecolor{blus}{cmyk}{1,1,0,0.6}
\definecolor{bluc}{cmyk}{1,1,0,0.1}
\definecolor{verde}{cmyk}{0.92,0,0.59,0.25}
\definecolor{verdec}{cmyk}{0.92,0,0.59,0.15}
\definecolor{verdes}{cmyk}{0.92,0,0.59,0.4}

 \def\be   {\begin{equation}}   \def\ee   {\end{equation}}
 \def\ba   {\begin{array}}      \def\ea   {\end{array}}

\font\tenrsfs=rsfs10 at 12pt
\font\sevenrsfs=rsfs7
\font\fiversfs=rsfs5
\newfam\rsfsfam
\textfont\rsfsfam=\tenrsfs
\scriptfont\rsfsfam=\sevenrsfs
\scriptscriptfont\rsfsfam=\fiversfs
\def\mathscr#1{{\fam\rsfsfam\relax#1}}

\def\circa#1{\,\raise.3ex\hbox{$#1$\kern-.75em\lower1ex\hbox{$\sim$}}\,}

\newcommand{\beq}{\begin{equation}}
\newcommand{\eeq}{\end{equation}}

\font\ital=cmu10 

\makeatletter
\def\hhref#1{\href{http://arxiv.org/abs/#1}{arXiv:#1}}
\usepackage{xstring} 
\newcommand{\hhrefq}[1]{\IfSubStr{#1}{:}{\href{http://inspirehep.net/search?ln=en&ln=en&p=#1&of=hb&action_search=Search&sf=&so=d&rm=&rg=25&sc=0}{InSpires:#1}}{\hhref{#1}}}

\def\art{\@ifnextchar[{\eart}{\oart}}
\def\eart[#1]#2#3#4#5#6{{\rm #2}, {\em #3 \bf #4} {\rm (#6) #5} ({\em #1})}
\def\article{\@ifnextchar[{\earticle}{\oarticle}}
\def\oarticle#1#2#3#4#5#6{{\rm #1}, {\ital ``#6''}, {\rm #2 #3 (#5) #4}}
\def\earticle[#1]#2#3#4#5#6#7{{\rm #2}, {\ital ``#7''}, {\rm #3 #4 (#6) #5}  [\hhrefq{#1}]}
\def\hepart[#1]#2{{\rm #2, \ital#1}}
\def\heparticle[#1]#2#3{#2, {\ital ``#3''} [\hhrefq{#1}]}
\newcommand{\doi}[1]{\href{http://dx.doi.org/#1}{[link]}}

\def\hhref#1{\href{http://arxiv.org/abs/#1}{arXiv:#1}} 

\begin{document}

\thispagestyle{empty}
\vspace{0.1cm}
\begin{center}
{\huge \bf \color{rossos}   
Dark Matter from freeze-in \\[1ex]  and its inhomogeneities}  \\[4ex]
{\bf\large Alessandro Strumia}  \\[2ex]
{\it Dipartimento di Fisica dell'Universit{\`a} di Pisa, Italia}\\[4ex]
{\large\bf\color{blus} Abstract}
\begin{quote}\large
We consider generic freeze-in processes for generation of Dark Matter,
together with the consequent re-thermalization of the Standard Model fluid.
We find that Dark Matter inherits the Standard Model adiabatic inhomogeneities
on the cosmological scales probed by current observations, that were
super-horizon during freeze-in.
Thereby, freeze-in satisfies the bounds on iso-curvature perturbations.
\end{quote}
\end{center}

\tableofcontents

\section{Introduction}

Freeze-in is a possible mechanism that could have generated the Dark Matter (DM) cosmological abundance~\cite{freeze-in}.
It assumes that the Standard Model (SM) cosmological thermal plasma was not initially accompanied by any DM abundance.
Since all SM components self-interact thermalising to a common temperature,
cosmological inhomogeneities were initially adiabatic.

Next, `freeze-in' particle physics processes produce DM particles with mass $M$ out of the SM plasma.
For example, one can have  decays SM $\to$ DM DM or scatterings SM SM $\to$ DM DM,
dominated either at large temperatures $T\gg M$ (`UV-dominated freeze-in')
or at low temperatures $T \sim M$ (`IR-dominated freeze-in).
In order to match the observed cosmological  DM density~\cite{Planckresults},
the rate of freeze-in processes must be much smaller than the Hubble rate $H$.
Freeze-in automatically generates DM inhomogeneities out of SM inhomogeneities.

Observations are consistent with dominant adiabatic inhomogeneities
(namely, the SM/DM fluid is the same everywhere),
while iso-curvature inhomogeneities 
(namely, DM inhomogeneities different from SM inhomogeneities)
are constrained, on cosmological scales, to be below a few $\%$ level~\cite{Planckresults}.

We consider if freeze-in leads to acceptable DM inhomogeneities.

\smallskip

Weinberg answered positively this issue for thermal freeze-out:
since freeze-out dominantly happens in the non-relativistic regime, 
computing inhomogeneities in the DM number density was enough~\cite{astro-ph/0405397}.
On the other hand, freeze-in can be relativistic, and the
 iso-curvature issue started being considered recently:
\cite{Bellomo}  claims that a specific freeze-in model is excluded because it generates too large scale-independent
iso-curvature perturbations.
The authors of \cite{Bellomo} argue that all freeze-in models are similarly problematic.
In the model considered in~\cite{Bellomo} DM has a small electric charge and is thereby produced by 
IR-dominated scatterings of two SM particles, such as $e^-e^+\to {\rm DM}\,{\rm DM}$.
This generates, at any given time, a contribution to the DM density $\rho_{\rm DM}$
proportional to the square of the SM density $\rho_{\rm SM}$, 
and DM inhomogeneities might be not proportional to SM inhomogeneities.
However, one must consider the cumulative cosmological process taking into account
that all regions of the Universe undergo a similarly diluting $\rho_{\rm SM}$.
As we will see, this leads to negligible iso-curvature effects.
A simple argument is presented in section~\ref{separate}, and the
general formalism is used in section~\ref{general}.
Section~\ref{concl} presents our conclusions.

\section{Intuitive argument based on `separate universes'}\label{separate}
We start presenting an intuitive argument.
Working in the Newtonian gauge, 
\beq ds^2  = -[1+2\Phi(t,\vec x)]dt^2+a^2(t) [1-2\Psi(t,\vec x)] d\vec{x}^2\eeq
the primordial adiabatic perturbations $\delta\rho_\alpha(t,\vec x)$
in the density $\rho_\alpha(t)$ of a fluid $\alpha$
can be characterised in a simple geometric way as~\cite{astro-ph/0003278,astro-ph/0306498,astro-ph/0401313,astro-ph/0405397}
\beq\label{eq:deltat}
\delta \rho _\alpha =  \frac{d\rho_\alpha}{dt}  \, \delta t
\eeq
working at first-order in the small $\delta\rho_\alpha\ll \rho_\alpha$.
In eq.\eq{deltat}  $\delta t (t,\vec x)$ is some universal function common to all fluids
that can be intuitively thought as a delay in the time evolution of the different regions.

\smallskip

Observation constrain iso-curvature perturbations only on scales comparable to the horizon today,
while the freeze-in DM density was generated before matter/radiation equality (much before in most freeze-in models).
This means that  we only need to worry if freeze-in generated
iso-curvature perturbations on scales much larger than the small horizon at freeze-in.

We can thus apply the  `separate universes' picture (see e.g.~\cite{astro-ph/0003278}):
the very early Universe at freeze-in can be thought as many homogeneous regions without causal contact,
given that inhomogeneities on different scale evolve independently in first-order approximation.
Freeze-in dynamics produces DM with adiabatic perturbations because
all regions  undergo the same dynamics, up to the delay $\delta t$.
So eq.\eq{deltat} holds for the DM density, no matter how complicated the freeze-in dynamics is.
Explicitly, the Boltzmann equation 
for the homogeneous small DM number density  $n_{\rm DM}$
is $ d(n_{\rm DM}/s)/d\ln T \simeq \gamma/Hs$, where $s$ is the entropy density,
$H$ is the Hubble rate, and
$\gamma (T)$ is the space-time density rate of freeze-in processes that produce one DM particle 
out of the SM plasma at temperature $T$.
Integrating this equation leads to 
\beq \label{eq:FIestimate}
 \frac{{n}_{\rm DM}}{s} = \int \frac{d T}{ T}   \frac{  \gamma ( T)}{ H  s}.
\eeq
Interpreting eq.\eq{FIestimate} in the `separate universes' picture implies that,
in regions where the SM plasma was denser,
freeze-in initially produced more DM by some amount that depends on the freeze-in model,
but in these region the DM average density changed more rapidly leading to adiabatic DM inhomogeneities.
The above discussion explicitly verifies how, in the special freeze-in case, 
the `separate universes' regions undergo the same evolution, up to the time delay.

\smallskip

The next section substantiates the above intuitive reasoning by explicit computations.




\section{Iso-curvature perturbations during freeze-in?}\label{general}
A general formalism 
to compute the cosmological evolution of inhomogeneities in interacting fluids was developed in~\cite{Bardeen:1980kt,Kodama:1984ziu}.
We adopt its presentation as summarized in~\cite{astro-ph/0411703},
that makes more explicit the sources of iso-curvature inhomogeneities.

Simple first-order evolution equations for the various densities
are obtained by combining the Einstein gravity equations into
the conservation of the energy-momentum  tensor $T^{\mu\nu} = \sum_\alpha T^{\mu\nu}_{(\alpha )}$.
The energy-momentum  tensor $T^{\mu\nu}_{(\alpha )}$
of fluid $\alpha $ only  is not conserved because interactions transfer
energy-momentum $Q^\nu_{(\alpha)}$ to other fluids. So one has
\beq \nabla_\mu T^{\mu\nu}_{(\alpha )} = Q^\nu_{(\alpha)}\qquad
\hbox{with}\qquad  \sum_\alpha Q^\nu_{(\alpha)}=0\eeq 
because of total energy conservation.
In the homogeneous limit, this implies that the average densities evolve as
$\dot\rho_\alpha + 3H (\rho_\alpha+\wp_\alpha)=Q^0_{(\alpha)} \equiv Q_\alpha  $,
where $\wp_\alpha$ is the pressure of fluid $\alpha$.
The energy component of $Q_{(\alpha)}$
is expanded in small inhomogeneities as
$Q_{(\alpha)0}=-Q_\alpha (1+\Phi) - \delta Q_\alpha$~\cite{astro-ph/0411703}
so that $\sum_\alpha \delta Q_\alpha=0$ by total energy conservation.
The total density is $\rho=\sum_\alpha \rho_\alpha$.

It is useful to write equations in terms of the curvature perturbation
$\zeta=-H [ \Psi/H+\delta\rho/\dot\rho]$,
which is the relative displacement between uniform-density and uniform-curvature surfaces.
This curvature perturbation can be defined for each fluid
\beq \zeta_\alpha=-\Psi - H \frac{\delta \rho_\alpha}{\dot\rho_\alpha}\eeq
and it evolves as~\cite{astro-ph/0411703}
\beq\label{eq:dzeta}
\dot\zeta_\alpha = - \frac{H}{\dot\rho_\alpha} \delta Q_{{\rm intr},\alpha}+
3\frac{H^2}{\dot\rho_\alpha}\delta \wp_{{\rm intr},\alpha} - \dot H\frac{Q_\alpha}{\dot\rho_\alpha}
\left(\frac{\delta\rho_\alpha}{\dot\rho_\alpha} - \frac{\delta\rho}{\rho}\right)+{\cal O}(k^2)\eeq
where 
$ \delta Q_{{\rm intr},\alpha}$ and $\delta \wp_{{\rm intr},\alpha} $ will be defined later.
As usual, small perturbations are conveniently expanded in comoving Fourier modes $k$,
and the `separate universe' argument amounts to consider the limit $k\to 0$ of the full equations.
We focus on large super-horizon scales, thereby omitting the label $k$ and
neglecting Laplacians and other terms suppressed by $k^2/a^2 H^2$.
Such terms are indeed negligible whenever freeze-in occurs way before matter/radiation equality,
for relevant cosmological scales $k$.

\smallskip

The equations\eq{dzeta} can be written in a slightly more convenient form by avoiding
using the total density $\rho$ and defining instead the
iso-curvature relative perturbations $S_{\alpha\beta}$ between two fluids $\alpha$ and $\beta$
\beq S_{\alpha\beta}\equiv 3(\zeta_\alpha-\zeta_\beta)=-3H\left(\frac{\delta \rho_\alpha}{\dot\rho_\alpha}-
\frac{\delta\rho_\beta}{\dot\rho_\beta}\right)
\eeq 
that evolve as
\beq \label{eq:dS}
\dot S_{\alpha\beta} =-3H\left(
\frac{\delta Q_{{\rm intr},\alpha}-3H \, \delta \wp_{{\rm intr},\alpha}}{\dot\rho_\alpha}-
\frac{\delta Q_{{\rm intr},\beta} - 3H\,\delta \wp_{{\rm intr},\beta}}{\dot\rho_\beta} \right)+\dot S_{\alpha\beta}^{\rm mul}+{\cal O}(k^2).
\eeq
 We again ignore the terms suppressed by $k^2$.  We can also ignore the `multiplicative' terms
(namely, those proportional to combinations of $S_{\alpha'\beta'}$ terms)~\cite{astro-ph/0411703}
{\beq  \dot S_{\alpha\beta}^{\rm mul}=
\frac{\dot H}{2H}\left[\left(\frac{Q_\alpha}{\dot\rho_\alpha} + \frac{Q_\beta}{\dot\rho_\beta} \right)S_{\alpha\beta}
+\left(\frac{Q_\alpha}{\dot\rho_\alpha} - \frac{Q_\beta}{\dot\rho_\beta} \right)\sum_\gamma \frac{\dot\rho_\gamma}{\dot\rho}
(S_{\alpha\gamma}+S_{\beta\gamma})\right]\eeq
 because} we are only concerned in understanding if 
non-zero iso-curvature perturbations
are generated by the `source' terms explicitly shown in eq.\eq{dS}.
The formalism summarized in~\cite{astro-ph/0411703} makes clear that,
in the long-wavelength limit $k\to 0$,
iso-curvature perturbations are only sourced by the
non-adiabatic energy transfer $ \delta Q_{{\rm intr},\alpha}$ and by the
non-adiabatic pressure $\delta \wp_{{\rm intr},\alpha} $ intrinsic in each fluid $\alpha$.
These terms will be now be defined and evaluated.


\subsection{Intrinsic non-adiabatic energy transfer}\label{Enad}
One source of iso-curvature perturbations is the 
intrinsic non-adiabatic energy transfer, 
the part of energy transfer $\delta Q_\alpha$ from fluid $\alpha$
`biased' with respect to its energy density $\rho_\alpha$~\cite{astro-ph/0411703}:
\beq \delta Q_{{\rm intr},\alpha} \equiv \delta Q_\alpha - \frac{\dot Q_\alpha}{\dot\rho_\alpha} \delta \rho_\alpha.\eeq
We next consider its value during freeze-in, where the relevant fluids are
$\alpha=\{{\rm SM},{\rm DM}\}$.

\smallskip

The rate  of freeze-in particle collisions can be computed, in any given particle-physics model, as a function of the local 
temperature of the SM fluid, that also controls its density. 
Thereby the energy transfer  from the SM fluid only depends on its local density, 
$Q_{\rm SM}(\rho_{\rm SM})$.
Consequently 
$ \delta Q_{{\rm intr,SM}} = \delta Q_{\rm SM} - \delta \rho_{\rm SM} d Q_{\rm SM}/d\rho_{\rm SM}=0$ vanishes
in a generic freeze-in model.

\smallskip

Next, energy conservation demands $\delta Q_{\rm SM}+\delta Q_{\rm DM}=0$, so that
the intrinsic non-adiabatic energy transfer to the DM fluid can be written as
\beq \delta Q_{\rm intr,DM}\equiv
 \delta Q_{\rm DM}  - \frac{\dot Q_{\rm DM} }{\dot\rho_{\rm DM} } \delta \rho_{\rm DM} 
= \dot Q_{\rm DM} \left( \frac{\delta \rho_{\rm SM}}{\dot\rho_{\rm SM}}-
 \frac{\delta \rho_{\rm DM}}{\dot\rho_{\rm DM}}\right).
\eeq
This potential `source' terms thereby becomes a `multiplicative' term,
proportional to the relative entropy $S_{\rm SM,DM}$.
Since this is assumed to be initially vanishing, $ \delta Q_{\rm intr,DM}$ generates no isocurvature perturbation.

\subsection{Intrinsic non-adiabatic pressure}
The second kind of source term,
the non-adiabatic part of the pressure perturbation intrinsic of each fluid $\alpha$, is given by~\cite{astro-ph/0411703}
\beq \delta \wp_{{\rm intr},\alpha} = \delta \wp_\alpha - c_\alpha^2 \delta\rho_\alpha\qquad\hbox{where}\qquad
c_\alpha^2 = \dot \wp_\alpha/\dot\rho_\alpha
\eeq
is its adiabatic speed of sound.
This term 
vanishes when the pressure and energy inhomogeneities
respect the equation of state of the fluid, $\wp_\alpha(\rho_\alpha)$.

\smallskip


Freeze-in particle-physics processes 
contribute as $\delta \wp_{{\rm intr,SM}}\neq 0$ because they
convert SM particles into DM particles,
thereby inducing an energy and momentum loss of the SM fluid,
as dictated by  the specific freeze-in interaction,
that generically does not follow the equation of state of the SM fluid. 

As a simple example of this unbalance, freeze-in via the decay into DM particles of some SM particle
(or, in SM extensions, of some speculative new-physics particle tightly coupled to the SM) transfers more energy than pressure
($\dot\rho_{\rm SM}/\dot \wp_{\rm SM} > \rho_{\rm SM}/\wp_{\rm SM}$)
because {the decaying particles must be massive and thereby they decay slower when they 
have higher relativistic energy}.
{An unbalance also generically occurs in freeze-in scatterings, described by a cross-section 
$\sigma($SM SM $\to$ DM DM$)$ that only depends on the invariant energy $\sqrt{s}$
at leading order in the couplings (the motion with respect to the plasma enters at higher orders). 
The sign of $\delta \wp_{{\rm intr,SM}}$ is not fixed, as}
the energy dependence of $\sigma$  can either result in a larger energy transfer when the colliding 
SM particles have higher energy $E\circa{>}T$ 
(this can happen in UV-dominated freeze-in, via non-renormalizable interactions, for example gravitational~\cite{GravDM}) 
or when the colliding
SM particles have lower energy $E\circa{<} T$ (this can happen in IR-dominated freeze-in, via renormalizable interactions).
{As a possibly relevant special case}, $\delta \wp_{{\rm intr,SM}}$ is nearly-vanishing in
freeze-in models that only lead to the disappearance of ultra-relativistic SM particles,
as they (on angular average) satisfy the same  equation of state $\wp=\rho/3$ as the radiation-dominated SM fluid.

\smallskip

However, the fact that freeze-in {processes (decays and scatterings)} can contribute as $\delta \wp_{{\rm intr,SM}}\neq 0$ is
inconsequential, as we must also take into account the self-interactions of the SM fluid.
A multitude of SM particle processes allow the SM fluid to locally re-thermalize to its equation of state
with rates $\Gamma$ much faster than the Hubble rate and than the freeze-in rate.
Typically $\Gamma \sim g^2 T$  where $g\sim 1$ is a typical SM coupling, {such as a gauge coupling}.
The re-thermalizion processes conserve the SM energy $\rho_{\rm SM}$:
$\delta Q_{\rm SM}$ remains given by freeze-in processes only, so that
$ \delta Q_{{\rm intr,SM}}=0$ remains as in section~\ref{Enad}.
On the other hand, the SM pressure $\wp_{\rm SM}$ changes such that
the combination of the two processes (freeze-in and re-thermalization) leads to $\delta \wp_{{\rm intr,SM}}= 0$.

\medskip

This leaves $\delta \wp_{{\rm intr,DM}}$ as a possible source of iso-curvatures.
A self-thermalization argument parallel to what {just} discussed for the SM plasma implies $\delta \wp_{{\rm intr,DM}}=0$
if DM has significant self-interactions just after being produced during freeze-in.
This happens, for example, if DM is a multiplet under a dark gauge group~\cite{2012.12087}
that confines at a scale $\Lambda$ and if freeze-in happens at $T \gg \Lambda$.
If instead DM self-interactions are negligible, a formalism extended to higher moments may be needed,
but the physics is simple: DM particles free stream on sub-horizon scales, but not on large scales $k\to 0$.
The non-thermal DM distribution
$f(\vec x,t,q)=f_0(q) + \delta f (\vec x,t,q)$ produced by freeze-in  redshifts with scale factor $a$ as~\cite{astro-ph/9506072}
\beq
 \rho_{\rm DM} = \frac{1}{a^4} \int \frac{d^3q}{(2\pi)^3}E \,  f,\qquad
 \wp_{\rm DM} = \frac{1}{a^4} \int \frac{d^3q}{(2\pi)^3}\frac{q^2}{3E} \,  f  \eeq
where $q$ and $E=\sqrt{q^2 + a^2 M^2}$ are the comoving momentum and energy
{of the DM particle with mass $M$}.
Two limits are of special interest.
If freeze-in is IR-dominated, DM is only mildly relativistic, 
so that DM motion is soon red-shifted down to negligible pressure, $\wp_{\rm DM}\ll \rho_{\rm DM}$.
UV-dominated freeze-in can produce ultra-relativistic DM with $\wp_{\rm DM}/\rho_{\rm DM} \simeq \dot\wp_{\rm DM}/\dot\rho_{\rm DM}  \simeq 1/3$,
that becomes non-relativistic only later when the SM cools down to
temperatures comparable to the DM mass $M$, while the horizon reaches larger scales.

\section{Conclusions}\label{concl}
We considered generic models of freeze-in
(from decays, from scatterings, IR-dominated, UV-dominated...)
finding that the generated Dark Matter inherits the Standard Model adiabatic inhomogeneities
on the cosmological scales probed by current observations, that were
super-horizon during freeze-in.
In section~\ref{separate} we presented an intuitive argument based on the well-known `separate universe' picture.
This was substantiated in section~\ref{general} by checking the explicit sources of iso-curvature perturbations
on super-horizon scales.

\smallskip

Iso-curvature perturbations can only be generated on small scales that were sub-horizon during freeze-in:
this effect can perhaps be relevant in models
where freeze-in happens at the lowest possible temperature $T\sim M \sim {\rm keV}$, 
possibly in the presence of 
dark long-range forces.

\smallskip

In conclusion, freeze-in appears a viable mechanism for generation of
the cosmological DM abundance.
Similar arguments hold for other particle-physics mechanisms 
such as `cannibalism'~\cite{cannibal} or `freeze-out and decay'.
Furthermore, baryogenesis mechanisms that involve elements similar to freeze-in
(such as leptogenesis from right-handed neutrinos with initially negligible abundance)
are similarly compatible with iso-curvature bounds.

\small

\paragraph{Acknowledgements}
We thank Guido d'Amico, Alessio Notari, Paolo Panci,
Michele Redi, Andrea Tesi for discussions.
This work was supported by the MIUR grant PRIN 2017FMJFMW.

\end{document}